\documentclass[a4paper,11pt]{article}
\usepackage[utf8]{inputenc}
\usepackage{mathtools,amsmath,amssymb}
\usepackage{graphicx}
\usepackage{lmodern}
\usepackage[T1]{fontenc}
\usepackage{microtype}
\usepackage[pagebackref=false]{hyperref}
\usepackage{rotate}

\usepackage[textsize=tiny]{todonotes}

\usepackage{lscape}

\usepackage{xspace}
\usepackage{bibentry}
\usepackage{easybmat}

\usepackage{graphicx}

\usepackage{subcaption}

\usepackage[numbers,sort&compress]{natbib}
\usepackage{hypernat}

\usepackage{tikz}
\usetikzlibrary{decorations.pathreplacing}
\numberwithin{equation}{section}

\usepackage[margin=2.5cm]{geometry}

\newcommand{\ID}{I}

\usepackage{enumitem}
\usepackage[misc]{ifsym}

\DeclareMathOperator{\tr}{tr}

\DeclareMathOperator{\PP}{P}

\newcommand{\calA}{\mathcal{A}}
\newcommand{\calB}{\mathcal{B}}
\newcommand{\calC}{\mathcal{C}}
\newcommand{\calD}{\mathcal{D}}
\newcommand{\Gop}{\mathcal{G}}

\newcommand{\qa}{1}
\newcommand{\qb}{q}
\newcommand{\pa}{1}
\newcommand{\pb}{p}

\newcommand{\hasse}{
\tikzstyle{boxi} = [
    rectangle,
    node distance=1.3 cm,
    text width=3.5 em,
    text centered,
    minimum height=1.3 em,
    minimum width=2 cm
]
\tikzstyle{container} = [
    rectangle,
    draw,
    inner sep=3cm,
    dashed
]

\tikzstyle{empty} = [
]

\tikzstyle{line} = [
    draw,
    fill=black!100,
->,
    thick,
    black!80
]

\begin{tikzpicture}

    \node [empty](origin){};
    
    \node [boxi, below=of origin](q0){$|(1,2,3)\rangle$};
    \node [boxi, above right=1.5cm of q0] (q3) {$|(2,3)\rangle$};
    \node [boxi, above =1.5cm of q0] (q2) {$|(1,3)\rangle$};
    \node [boxi, above left=1.5cm of q0] (q1) {$|(1,2)\rangle$};

    \node [boxi, above =1.5cm of q1] (q12) {$|(1)\rangle$};
    \node [boxi, above right =1.5cm of q1] (q13) {$|(2)\rangle$};
    \node [boxi, above =1.5cm of q3] (q23) {$|(3)\rangle$};

    \node [boxi, above right =1.5cm of q12] (q123) {$|I_0=\emptyset\rangle$};
    
    \path [line] (q3) -- (q0);
    \path [line] (q2) --  (q0);
    \path [line] (q1) -- (q0);
    
    \path [line] (q12) -- (q1);
    \path [line] (q23) --  (q2);
    \path [line] (q13) -- (q3);
    
    \path [line] (q12) -- (q2);
    \path [line] (q23) --  (q3);
    \path [line] (q13) -- (q1);

    \path [line] (q123) -- (q12);
    \path [line] (q123) --  (q13);
    \path [line] (q123) -- (q23);
    
    \node [boxi, below right =1cm of q2] {$\mathfrak{F}(I_h;i)$};

\end{tikzpicture}
}

\begin{document}

\begingroup\parindent0pt
\vspace*{4em}
\centering
\begingroup\Large\bf

Eigenstates of triangularisable open XXX spin chains and closed-form solutions for the steady state of the open SSEP

\par\endgroup
\vspace{5em}
\begingroup\large
{ Rouven Frassek}\\
\par\endgroup
\vspace{1em}
\begingroup
Max-Planck-Institut für Mathematik, \\
   Vivatsgasse 7, 53111 Bonn, Germany\\
\par\endgroup
\vspace{1em}

\vspace{5em}

\vfill
\begin{abstract}
\noindent
In this article we study the relation between the eigenstates of open rational spin~$\frac{1}{2}$ Heisenberg chains with different boundary conditions. The focus lies on the relation between the spin chain with diagonal boundary conditions and the spin chain with triangular boundary conditions as well as the class of  spin chains that can be brought to such form by certain similarity transformations in the physical space. The boundary driven  Symmetric Simple Exclusion Process (open SSEP) belongs to the latter.
We derive a transformation that maps the eigenvectors of the diagonal spin chain to the eigenvectors of the triangular chain. 
This transformation yields an essential simplification for determining the states beyond half-filling. It allows to first determine the eigenstates of the diagonal chain through the Bethe ansatz on the fully excited reference state and subsequently map them to the triangular chain for which only the vacuum serves as a reference state. 
In particular the transformed reference state, i.e. the fully excited  eigenstate of the triangular chain,  is presented at any length of the chain. 
It can be mapped to the steady state of the open SSEP. This results in a concise closed-form expression for the probabilities of particle distributions and correlation functions in the steady state. Further, the complete set of eigenstates of the Markov generator is expressed in terms of the eigenstates of the diagonal open chain.
\end{abstract}
\vfill
\endgroup
\thispagestyle{empty}
\newpage
%
\setcounter{tocdepth}{2}

\tableofcontents

\section{Introduction}\label{sec:intro}

Since Sklyanin's formulation of the quantum inverse scattering method for integrable open spin chains 
\cite{Sklyanin:1988yz} there is a precise recipe to construct local  Hamiltonians  and transfer matrices of such models. The construction is based on solutions to the Yang-Baxter equation (R-matrices) as well as solutions to the boundary Yang-Baxter equation (K-matrices) and naturally generalises the quantum inverse scattering method for closed spin chains, see e.g.~\cite{Faddeev:1996iy} for an overview.
The framework allows to solve integrable models using the algebraic Bethe ansatz; the algebraic version of the coordinate Bethe ansatz which goes back to \cite{Bethe1931} for the closed  and to \cite{gaudin1983fonction,Alcaraz:1987uk} for the open Heisenberg chain.

The ordinary algebraic or coordinate Bethe ansatz is applicable if the magnon number of the spin chain is a  conserved quantity. In particular this is the case for the open Heisenberg spin chain with diagonal boundary conditions. In this case the magnons (excitations) of the chain are the up-spins in the sea of down-spins or vice versa. However, if this symmetry is broken the algebraic Bethe ansatz has to be modified. Here we are dealing with the rather well studied case of triangular boundary conditions. The eigensystem of such open chains was studied  within the algebraic Bethe ansatz in \cite{Melo:2005gut,BCR}, see also  \cite{Antonio:2014qxa}. For the coordinate Bethe ansatz we refer the reader to \cite{Crampe:2011fm}. The Hamiltonian of the triangular chain that we study in this article is of the form 
\begin{equation}\label{eq:triham}
H_\Delta=
p\,\sigma_3^{[1]}+\Delta\,\sigma_+^{[1]}+
 \sum_{k=1}^{N-1}\left(\vec \sigma^{[k]}\cdot\vec\sigma^{[k+1]}-\ID\right)+q\,\sigma_3^{[N]}\,.
\end{equation} 
Here $\sigma_{3,\pm}^{[i]}$ denote the Pauli matrices with $\sigma_\pm=\frac{1}{2}(\sigma_1\pm i\sigma_2)$ at site $i$ while $\pb, \qb,\Delta$ are complex parameters and $\ID$ denotes the identity matrix. If $\Delta=0$ we recover the Hamiltonian of the diagonal chain. We remark that the triangular spin chain is governed by a homogeneous Baxter equation which is the same as for the diagonal case. As a consequence the energy spectra coincide, however, the eigenvectors differ for the two types of spin chains!

The class of spin chains that can be solved using these methods does not only include the open spin chains with triangular boundaries but also all spin chains that can be brought to such form, i.e. that are triangularisable, by certain similarity transformations in the quantum space. 
More precisely, all integrable spin chains with a nearest neighbor Hamiltonian $\tilde H_\Delta$ that is related to $H_\Delta$ via 
\begin{equation}\label{eq:univham2}
H_\Delta=c_1 \, \mathcal{S}_G^{-1}\tilde H_\Delta\,\mathcal{S}_G+c_0\,,
\end{equation} 
belong to this class. 
Here  $c_1$ and $c_0$ are complex parameters. Further, the similarity transformation in the $2^N$-dimensional quantum (physical) space denotes the $N$-fold tensor product of a non-degenerate $2\times 2$ matrix
\begin{equation}\label{eq:simtr}
 \mathcal{S}_G=G\otimes G\otimes\ldots\otimes G\,.
\end{equation} 
Such transformation may change the boundary terms of the Hamiltonian but leaves the bulk terms  invariant. 
Obviously the energy spectrum or equivalently the Baxter equation is unchanged by such similarity transformation. The eigenvectors are obtained from the ones of the triangular chain by multiplying them with the transformation matrix $ \mathcal{S}_G$ in \eqref{eq:simtr}. The relations between the spectrum and eigenvectors of the spin chains discussed here are summarised in Figure~\ref{fig:tab}. We notice already at this stage that our table would be more uniform if there was a relation between the eigenstates of the diagonal and triangular chain which is similar to the relation between the eigenstates  of the triangular and triangularisable chains. Such relation between $|\psi_m^0\rangle$ and $|\psi_m^\Delta\rangle$ will be presented in the next Section~\ref{sec:tri}.

\begin{figure}
 \begin{center}
\renewcommand{\arraystretch}{1.5}
\begin{tabular}{l| c| c| c}
 &diagonal chain&triangular chain&triangularisable chains\\
 \hline
 Hamiltonian & $H_0$ & $H_\Delta$ & $\tilde H_\Delta$ \\ 
 \hline
 eigenvalues & $E_m^0$ & $E_m^\Delta=E_m^0$&$\tilde E_m^\Delta=\left(E_m^0-c_0\right)/c_1$ \\  
 \hline
 eigenstates & $|\psi^0_m\rangle$ & $|\psi^\Delta_m\rangle$ &   $|\tilde \psi_m^\Delta\rangle=\mathcal{S}_G|\psi_m^\Delta\rangle$
\end{tabular}
\renewcommand{\arraystretch}{1}
\end{center}
\caption{Overview of the energy spectrum and eigenvectors of the open spin chain Hamiltonians with diagonal and triangular boundaries as well as the class of chains that can be brought to a triangular form by a similarity transformation of type $\mathcal{S}_G$.}
\label{fig:tab}
\end{figure}

A prominent example of a spin chain that can be brought to a triangular form is the one underlying the Symmetric Simple Exclusion Process (SSEP) with reservoirs, see e.g. \cite{schutz2000exactly,Mallick} for an overview. Of particular interest in these types of non-equilibrium processes are its properties in the steady state, i.e. the state which remains steady in time when evolved by the Markov generator. For an introduction we refer the reader to \cite{blythe2007nonequilibrium} and \cite{Derrida:1992vu}. The Markov generator of the open SSEP can be mapped to the Hamiltonian of an open Heisenberg spin chain whose boundary terms can be brought to a triangular form, see in particular 
\cite{Crampe:2014aoa}  for a presentation within the quantum inverse scattering method as well as \cite{PhysRevLett.95.240601,degier2006}.
The knowledge of the open spin chain spectrum has been used in \cite{degier2006} to compute the scaling behavior in length of the energy gap between the steady state and the first excited state.
The steady state of the open SSEP  can be computed using the DEHP matrix product ansatz \cite{Derrida:1992vu}  which however does not directly refer to the integrable structure of the process. Naturally one would expect that the steady state can be constructed using the Bethe ansatz in \cite{Melo:2005gut,BCR}. This is indeed the case but as observed in  \cite{Crampe:2014aoa} the steady state corresponds the state built from the maximal number of excitations on the vacuum reference state. The corresponding Bethe roots are non-trivial and an analytic construction of the Bethe vector in terms of creation operators on the reference state seems out of reach. In fact, there is only one reference state known for the triangular open spin chain and not two as in the diagonal case. As we will see, the steady state corresponds to the other, fully excited, reference state.

In the following we provide a method which allows to construct the eigenstates of the open spin chain with triangular boundary conditions $|\psi_m^\Delta\rangle$ from the corresponding eigenstate of the open spin chain with diagonal boundaries $|\psi_m^0\rangle$, cf.~Figure~\ref{fig:tab}.  Keeping in mind that in the case of the diagonal open chain the eigenstates $|\psi_m^0\rangle$ can be determined from two possible reference states we never have to solve Bethe  equation beyond half-filling to determine the eigenstates of the triangular chain.  As a consequence, the method is particularly powerful to study the fully excited reference state as there are no Bethe equations to be solved. We give an explicit formula for it and show how it relates to the steady state of the open SSEP. The representation found for the steady state turns out to be particuarly convenient to compute its components and correlation functions in closed form.

The paper is structured as follows. In Section~\ref{sec:base} we review some basics about the transfer matrix construction of open spin chains within the quantum inverse scattering method focusing on diagonal and triangular K-matrices. In  Section~\ref{sec:evs} we present and prove a relation between the eigenstates of the triangular and diagonal chain involving the elements of the monodromy matrix. The formula relating the eigenvectors can be further simplified by taking the limit of large spectral parameter. This is discussed in Section~\ref{sec:evslarge}. In Section~\ref{sec:grnd} we apply the method presented in Section~\ref{sec:tri} to derive a closed-form expression for an eigenvector of the triangular chain. This eigenvector corresponds to the other, fully excited,  reference state of the diagonal chain and is directly related to the steady state of the open SSEP. We review the open SSEP in Section~\ref{sec:ssepqism} and give the precise relation to the eigensystem of the triangular chain in Section~\ref{sec:ssepresuts}. This section also contains a closed form expression for the probabilities and correlation functions in the steady state. We conclude in Section~\ref{sec:conc}. The appendix is dedicated to the computation of the eigenvalue of the reference state (Appendix~\ref{app:fcr}), the algebraic Bethe ansatz for the diagonal chain (Appendix~\ref{app:bae}) and also  contains a brief summary of the DEHP matrix product ansatz for the open SSEP (Appendix~\ref{sec:dehp}).

\section{The triangular spin chain}\label{sec:tri}
In this section we study the relation between the eigenvectors of open Heisenberg spin chains with diagonal and triangular boundary conditions. First, in Section~\ref{sec:base}, we review the quantum inverse scattering method for open spin chains following \cite{Faddeev:1996iy,Sklyanin:1988yz}. We discuss how the Hamiltonian in \eqref{eq:triham} is obtained from the transfer matrix and write down Baxter's TQ equation along with the Bethe equations. In Section~\ref{sec:evs} we spell out the relation between the eigenvectors of the two types of spin chains and provide a proof of it. This relation \eqref{eq:vecrel} can be seen as the main result of the paper as most of what is discussed afterwards follows from it. In particular the relation \eqref{eq:vecrel2} obtained in Section~\ref{sec:evslarge} emerges in the limit of large spectral parameter where the action of the transfer matrix becomes local.

\subsection{Quantum inverse scattering method}\label{sec:base}
In this section we review some basics of the quantum inverse scattering method for open Heisenberg spin chains. We construct the transfer matrix, derive the Hamiltonian \eqref{eq:triham} and briefly discuss the Bethe ansatz.

The R-matrix of the Heisenberg XXX spin chain is well known. It is defined in terms of the spectral parameter $x\in \mathbb{C}$ and the permutation operator as
\begin{equation}\label{eq:rmat}
 R(x)=x+P\,,\qquad\text{with}\qquad P=\sum_{i,j=1}^2e_{ij}\otimes e_{ji}\,,
\end{equation} 
and satisfies the Yang-Baxter equation.
Here we introduced the elementary $2\times2$ matrices $e_{ij}$ defined via $\left(e_{ij}\right)_{kl}=\delta_{ik}\delta_{jl}$. They can be expressed in terms of the Pauli matrices introduced earlier and the $2\times2$ identity matrix. For our purposes it is often more convenient to work with the elementary matrices. The R-matrix in \eqref{eq:rmat} naturally acts on the tensor product of two spaces
$R_{a,b}(x)\in End(\mathbb{C}^2\otimes\mathbb{C}^2)$ denoted by $a$ and $b$. This allows to define the spin chain monodromies via 
\begin{equation}\label{eq:singmon}
 M_a(x)=R_{a,1}(x)\cdots R_{a,N}(x)\,,\qquad  \hat M_a(x)=R_{a,N}(x)\cdots R_{a,1}(x)\,.
\end{equation} 
Here the R-matrices $R_{a,i}(x)$ are multiplied in the common so-called auxiliary space $V_a=\mathbb{C}^2$ while the $N$-fold tensor product is taken in the so called quantum space $V=V_1\otimes\ldots\otimes V_N$ such that $R_{a,i}(x)$ acts non-trivially on $V_i$ which corresponds to $i$-th spin chain site.

After having introduced the single-row monodromies it is common to define the double-row monodromy which involves one K-matrix $\hat K(x)$ that acts trivially in the quantum space and non-trivially in the auxiliary space
\begin{equation}\label{eq:drow}
 U_a(x)=M_a(x)\hat K_a(x)\hat M_a(x)\,.
\end{equation} 
We focus on the case where the K-matrix $\hat K(x)$ is diagonal and of the form
\begin{equation}
  \hat K(x)=\left(\begin{array}{cc}
             \qa+\qb x&0\\
             0&\qa-\qb x 
            \end{array}
\right)\,,
\end{equation} 
with the boundary parameters $\qb\in \mathbb{C}$. As discussed below, we can always bring the K-matrix to such a diagonal form. The double-row monodromy defined in \eqref{eq:drow} then satisfies the boundary Yang-Baxter equation. Finally we can define the transfer matrix by taking the trace in the auxiliary space  of the product of the K-matrix $K(x;\Delta)$ and the double-row monodromy. It reads
\begin{equation}\label{eq:trans}
 T_{\Delta}(x)=\tr_a K_a(x;\Delta)U_a(x)\,,
\end{equation} 
where $K_a(x;\Delta)$ only act non-trivially in the auxiliary space and is defined by the upper triangular matrix
\begin{equation}
 K(x;\Delta)=\left(\begin{array}{cc}
             \pa+\pb(x+1)&\Delta(x+1)\\
             0&\pa -\pb(x+1)
            \end{array}
\right)\,.
\end{equation} 
Here $\pb$ and $\Delta$ are boundary parameters $\pb,\Delta\in \mathbb{C}$. 
When setting $\Delta=0$ we recover the transfer matrix for the diagonal chain $T_0(x)$.  
For later purposes it is convenient to write the triangular transfer matrix in terms of the diagonal one. We have
\begin{equation}\label{eq:transis}
\begin{split}
 T_\Delta(x)&=T_0(x)+\Delta(x+1)\, \calC(x)\,,
\end{split}
\end{equation} 
where ${\calC}(x)$ denotes the lower left entry of the double-row monodromy matrix in the $2\times2$ auxiliary space
\begin{equation}\label{eq:drowabcd}
 U_a(x)=
 \left(\begin{array}{cc}
             \calA(x)& \calB(x)\\
              {\calC}(x)& \calD(x)
            \end{array}
\right)\,.
\end{equation} 
Here, along with the matrix ${\calC}(x)$, we defined $\calA(x)$, $\calB(x)$ and $\calD(x)$ that act on the quantum space $V$.

The transfer matrix defined in \eqref{eq:trans} commutes with itself at different values of the spectral parameter $[T_\Delta(x),T_\Delta(y)]=0$ and thus generates a commuting family of matrices.
The nearest neighbor Hamiltonian of the open triangular spin chain \eqref{eq:triham} can be extracted by taking the logarithmic derivative of the transfer matrix at the point where the R-matrix becomes a permutation \cite{Sklyanin:1988yz}, i.e. $x=0$. We find
\begin{equation}\label{eq:hamtri}
 H_\Delta=\frac{\partial}{\partial x}\log T_\Delta(x)|_{x=0}-(2N-1)\ID\,,
\end{equation} 
with the identity matrix denoted by $I$.
The transfer matrix introduced in \eqref{eq:trans} can be diagonalised using the algebraic Bethe ansatz as presented in \cite{BCR}. The reference state is given by
\begin{equation}\label{eq:ref}
 |\psi_0^\Delta\rangle=\left(\begin{array}{c}
                   1\\0
                  \end{array}
\right)\otimes\ldots\otimes \left(\begin{array}{c}
                   1\\0
                  \end{array}
\right)\,,
\end{equation} 
and the eigenvalue equation for the transfer matrix \eqref{eq:trans} can be written as
\begin{equation}\label{eq:eveq}
 T_\Delta(x)|\psi_m^\Delta\rangle =\Lambda_m(x) |\psi_m^\Delta\rangle\,,
\end{equation} 
where $m=0,1,2,\ldots,N$. 

A few comments are in order:
\begin{enumerate}[label=(\roman*)]
 \item For $\Delta=0$ the variable $m$ denotes the magnon number, i.e. the number of excitations on the reference state \eqref{eq:ref}.
\item For $\Delta\neq0$ the eigenstates do not contain a fixed number of excitations. They are overlaps of states with $k$ excitations where $k=0,1,\ldots,m$. This follows from the block triangular structure of the transfer matrix, cf.~\cite{BCR}.
\item For every $m$ in \eqref{eq:eveq} we have $\binom{N}{m}$ eigenvectors and eigenvalues. We assume that they are all non-degenerate but suppress an extra index to distinguish them.
\end{enumerate}

The eigenvalues of the transfer matrix can be obtained from the algebraic Bethe ansatz and  are expressed in terms of the Baxter TQ equation. The latter reads
\begin{equation}\label{eq:baxter}
\begin{split}
 \Lambda_m(x)&=(\pa+x\pb)(\qa+x \qb)\frac{2}{2x+1}\left(x+1\right)^{2N+1}\frac{Q_m(x-1)}{Q_m(x)}\\
 &\quad+(\pa-(x+1)\pb)(\qa-(x+1)\qb)\frac{2}{2x+1}x^{2N+1}\frac{Q_m(x+1)}{Q_m(x)}\,.
 \end{split}
\end{equation} 
Here we introduced Baxter Q-functions $Q_m(x)$ written in terms of the Bethe roots $x_i$ as
\begin{equation}
 Q_m(x)=\prod_{i=1}^m (x-x_i)(x+x_i+1)\,.
\end{equation} 
For $\Lambda_m(x)$ to be an eigenvalue of the transfer matrix the Bethe roots have to satisfy the Bethe equations
\begin{equation}\label{eq:bae}
\frac{(\pa+x_i\pb)(\qa+x_i \qb)\left(x_i+1\right)^{2N}}{(\pa-(x_i+1)\pb)(\qa-(x_i+1)\qb)x_i^{2N}}=\prod_{k\neq i}\frac{(x_i-x_k+1)(x_i+x_k+2)}{(x_i-x_k-1)(x_i+x_k)}\,,
\end{equation} 
where $i=1,\ldots,m$. 
 We note that the eigenvalues $\Lambda_m(x)$ are independent of $\Delta$ and stress again that the Baxter equation \eqref{eq:baxter} for the triangular chain coincides with the Baxter equation for the diagonal chain. The eigenvectors however depend on $\Delta$ and differ from the eigenvalues of the diagonal chain.  In particular it is easy to show that the other reference state of the diagonal chain
\begin{equation}\label{eq:Sigma}
|\psi_N^0\rangle= \left(\begin{array}{c}
                   0\\1
                  \end{array}
\right)\otimes\ldots\otimes \left(\begin{array}{c}
                   0\\1
                  \end{array}
\right)\,,
\end{equation} 
is \emph{not} an eigenstate of the transfer matrix of the triangular chain $T_\Delta(x)$ if $\Delta\neq 0$. So far the Bethe ansatz can only be performed using the reference state $|\psi_0^\Delta\rangle$ in \eqref{eq:ref}. As a consequence, one has to solve the Bethe equations for $m=N$ in order to compute the eigenstate $|\psi_N^\Delta\rangle$, see also Appendix~\ref{app:bae} for further details. We compute this state explicitly in Section~\ref{sec:grnd} using a different approach outlined in the following subsections.

We further  remark that in \cite{Melo:2005gut} the matrix $K$ was diagonal and $\hat K$ triangular. This chain is related to our case by a similarity transformation $\mathcal{S}_G$. This can be shown using that the R-matrix commutes with the tensor product of any non-degenerate matrix $G$, i.e. $ [R(x),G\otimes G]=0$.
It follows that the eigenvalues of the transfer matrices coincide up to a reparametrisation of the boundary parameters and only the eigenvectors change as discussed in Section~\ref{sec:intro}.

\subsection{Eigenvector construction of the triangular chain}\label{sec:evs}
In the following we propose a formula to construct the eigenvectors of the triangular transfer matrix $T_\Delta$ from the eigenvectors  of the diagonal transfer matrix $T_0$ of equal eigenvalues, cf.~\eqref{eq:transis}. As mentioned in the introduction, one can determine the diagonal chain's eigenvectors using the Bethe ansatz on two different reference states. Once this is done, we can use the construction presented below to obtain the corresponding eigenvector of the triangular chain. In partilcular, this yields a simplification for the reference states beyond half-filling.

Let $|\psi_m^0\rangle$ be an eigenvector of the transfer matrix $T_0(x)$ of eigenvalue $\Lambda_m(x)$.  The eigenvector $|\psi_m^\Delta\rangle$  of the triangular  transfer matrix $T_\Delta(x)$ corresponding to the \emph{same} eigenvalue $\Lambda_m(x)$ can then be obtained  via
\begin{equation}\label{eq:vecrel}
 |\psi_m^\Delta\rangle=\sum_{k=0}^m \Delta^k\, {\Gop}^k_{m,\epsilon}(x)|\psi^0_m\rangle\,\bigg|_{\epsilon=0}\,,
\end{equation} 
where 
\begin{equation}\label{eq:gofx}
 {\Gop}_{m,\epsilon}(x)=(x+1)\left[\left(\Lambda_m(x)+\epsilon\right)\ID-T_0(x)\right]^{-1}{\calC}(x)\,.
\end{equation} 
Here we introduced the parameter $\epsilon$ in order to make the inverse of the block diagonal operator $\Lambda_m(x)-T_0(x)$ well defined. However since this operator never acts on $|\psi_m^0\rangle$ in \eqref{eq:vecrel} and the spectrum of $T_0$ is assumed to be  non-degenerate we can set $\epsilon=0$ in the end without any subtleties. Further we note that \eqref{eq:vecrel} only holds as long as ${\calC}(x)$ is non-zero. In particular, this excludes the point $x=0$ where the double-row  monodromy is proportional to the identity matrix and thus ${\calC}(0)$ vanishes.

The relation in \eqref{eq:vecrel} can be shown by acting with the transfer matrix $T_\Delta$ as written in \eqref{eq:transis} on the states $ |\psi_m^\Delta\rangle$. We find
\begin{equation}
\begin{split}
 T_\Delta(x)|\psi_m^\Delta\rangle&=\left(T_0(x)+\Delta(x+1)\, {\calC}(x)\right)|\psi_m^\Delta\rangle\\
 &=T_0(x)|\psi_m^0\rangle+\sum_{k=1}^{m}\Delta^k\left((x+1){\calC}(x){\Gop}^{k-1}_{m,\epsilon}+T_0(x){\Gop}^{k}_{m,\epsilon}\right)|\psi_m^0\rangle|_{\epsilon=0}\\
 &\qquad\qquad\quad\;\;\;+\Delta^{m+1}(x+1){\calC}(x){\Gop}^{m}_{m,\epsilon}|\psi_m^0\rangle|_{\epsilon=0}\\
  &=\Lambda_m(x)|\psi_m^0\rangle+\left(\Lambda_m(x)+\epsilon\right)\sum_{k=1}^{m}\Delta^k{\Gop}^{k}_{m,\epsilon}|\psi_m^0\rangle|_{\epsilon=0}\\
  &=\Lambda_m(x)|\psi^\Delta_m\rangle\,.
 \end{split}
\end{equation} 
Here we used the identity 
\begin{equation}
 \left[(\Lambda_m(x)+\epsilon)\ID-T_0(x)\right]{\Gop}^k_{m,\epsilon}(x)|\psi^0_m\rangle=(x+1){\calC}(x){\Gop}^{k-1}_{m,\epsilon}(x)|\psi^0_m\rangle\,,
\end{equation} 
which follows from the definition of ${\Gop}^{k}_{m,\epsilon}$ in \eqref{eq:gofx} 
and the relation
\begin{equation}
{\calC}(x){\Gop}^{m}_{m,\epsilon}(x)|\psi^0_m\rangle|_{\epsilon=0}=0\,.
\end{equation} 
The latter follows from the fact that the operator ${\calC}(x)$ lowers the magnon number while the transfer matrix $T_0$ is block diagonal and does not change the magnon number. This can be expressed in terms of the commutation relations
\begin{equation}\label{eq:Ccom}
 [T_0(x),e_{22}^{tot}]=0\,,\qquad  [{\calC}(x),e_{22}^{tot}]={\calC}(x)\,,
\end{equation} 
where $e^{tot}_{22}=\sum_{i=1}^N e_{22}^{[i]}$ measures the magnon number.
As as consequence  we find that ${\Gop}^{m}_{m,\epsilon}(x)|\psi^0_m\rangle|_{\epsilon=0}$ is proportional to the ground state $|\psi_0^\Delta\rangle$ in \eqref{eq:ref} since
 $e_{22}^{tot}\,{\Gop}^{m}_{m,\epsilon}(x)|\psi^0_m\rangle|_{\epsilon=0}=0$
and thus is annihilated by ${\calC}(x)$. This concludes the proof of \eqref{eq:vecrel}.

We remark that the eigenvalues $\Lambda_m(x)$ and eigenvectors $|\psi_m^0\rangle$ can be obtained explicitly using the standard Bethe ansatz. The algebraic Bethe ansatz is summarised in Appendix~\ref{app:bae}.
Further, we point out that the eigenvectors $|\psi^\Delta_m\rangle$ are independent of the spectral parameter. Consequently, when expanding in terms of the spectral parameter, it is sufficient to consider only the term proportional to the identity as all other terms will vanish.
Further we stress that the construction at this level only relies on the relation \eqref{eq:transis} and the explicit form the transfer matrices has not been used. We thus remark that a slightly modified equation should hold for the XXZ chain. Here the term proportional to $\Delta$ will be a trigonometric function of $x$, cf.~\cite{Pimenta:2013cua,Belliard:2014fsa} where the Bethe ansatz is discussed. 

In the next section we send the spectral parameter $x$ to infinity where the action of the operators $\calA(x)$, $\calB(x)$, ${\calC(x)}$ and $\calD(x)$ in the double-row monodromy becomes local.

\subsection{Large $x$ limit}\label{sec:evslarge}
In the previous subsection we have introduced a map from the eigenvectors of the diagonal chain to the eigenvectors of the triangular chain which depends on the spectral parameter $x$ although the final result does not, see \eqref{eq:vecrel}. In this subsection we take the limit $x\to\infty$ where the action of the operator ${\Gop}^k_{m,\epsilon}(x)$ becomes local and remove the parameter $\epsilon$.

In order to take the limit $\lim_{x\to\infty}{\Gop}^k_{m,\epsilon}(x)$ in \eqref{eq:vecrel} we take a closer look at the large $x$ expansion of the entries in the double-row monodromy. We begin with the diagonal part which appears in the diagonal transfer matrix in the denominator of ${\Gop}^k_{m,\epsilon}(x)$. In terms of the entries of the double-row monodromy the transfer matrix $T_0(x)$ can be written as 
\begin{equation}\label{eq:t0ab}
 T_0(x)=\calA(x)+\calD(x)+(x+1)\pb\left(\calA(x)-\calD(x)\right)\,.
\end{equation} 
The entries of the double-row monodromy $\calA(x)$ and $\calD(x)$ are polynomials of degree $2N+1$ in the spectral parameter 
\begin{equation}
 \calA(x)=\sum_{i=0}^{2N+1}\calA_{i}\,x^i\,,\qquad \calD(x)=\sum_{i=0}^{2N+1}\calD_{i}\,x^i\,.
\end{equation} 
For our purposes it is enough to compute the coefficients in $x$ for the first three leading orders.
At leading order $x^{2N+1}$ the operators $\calA(x)$ and $\calD(x)$ are proportional to the identity. We simply get
\begin{equation}
 \calA_{2N+1}=\qb\,,\qquad  \calD_{2N+1}=-\qb\,.
\end{equation} 
At sub-leading order the coefficients are diagonal but not proportional to the identity. We find the local form
\begin{equation}
 \calA_{2N}=\qa+2\qb\sum_{i=1}^Ne_{11}^{[i]}\,,\qquad  \calD_{2N}=\qa-2\qb\sum_{i=1}^Ne_{22}^{[i]}\,.
\end{equation} 
Finally, at sub-sub-leading order we obtain the bi-local expressions
\begin{equation}
 \calA_{2N-1}=2\sum_{i=1}^N e_{11}^{[i]}+2\qb\sum_{i\neq j}e_{11}^{[i]}e_{11}^{[j]}+\qb\sum_{i=1}^N\left(e_{11}^{[i]}-e_{22}^{[i]}\right)\,,
\end{equation} 
\begin{equation}
 \calD_{2N-1}=2\sum_{i=1}^N e_{22}^{[i]}-2\qb\sum_{i\neq j}e_{22}^{[i]}e_{22}^{[j]}+\qb\sum_{i=1}^N\left(e_{11}^{[i]}-e_{22}^{[i]}\right)\,.
\end{equation} 
We can now write down the large $x$ expansion of the transfer matrix $T_0$ as given in \eqref{eq:t0ab}. It reads
\begin{equation}
\begin{split}
 T_0(x)&=2\pb\qb x^{2N+2}+2\pb\qb(N+1)x^{2N+1}+
 \\&\quad+
 2 \left[(\pa + N \pb) (\qa + N \qb) + 2\pb\qb e_{22}^{tot}\left(\left(e_{22}^{tot}-N\right)- \frac{\qa}{\qb} - \frac{\pa}{\pb} \right) \right]x^{2N}+\ldots\,.
 \end{split}
\end{equation} 
Evidently the first two orders in the expansion of the diagonal transfer matrix are proportional to the identity matrix. At order $x^{2N}$ we observe a dependence on the the operator $e_{22}^{tot}$ introduced in Section~\ref{sec:evs} which measures the magnon number. The expansion of the diagonal transfer matrix immediately allows us to read of the expansion of its eigenvalues $\Lambda_m$ as $e_{22}^{tot}$ acts diagonally on the eigenvectors $e_{22}^{tot}|\psi_m^0\rangle=m|\psi_m^0\rangle$. Thus we conclude that at leading order in the spectral parameter $x$ the difference that appears in the denominator of ${\Gop}_{m,\epsilon}(x)$ takes the form
\begin{equation}\label{eq:ltexp}
\begin{split}
 \Lambda_m(x)-T_0(x)&=4 \pb \qb\left(m-e_{22}^{tot}\right) \left( \left(m-N+e_{22}^{tot}\right) - \frac{\qa}{\qb} - \frac{\pa}{\pb} \right)x^{2N}+\ldots\,.
   \end{split}
\end{equation} 

We now turn to the large $x$ expansion of ${\calC}(x)$. By definition ${\calC}(x)$ is a polynomial of degree $2N-1$ in the spectral parameter
\begin{equation}
{\calC}(x)=\sum_{i=0}^{2N-1}\calC_{i}\,x^i\,.
\end{equation} 
In this case it is enough to consider the leading order in $x$. By expanding the single-row monodromies we find the bi-local expression
\begin{equation}\label{eq:cexp}
 \calC_{2N-1}=2\sum_{i=1}^N e_{12}^{[i]}+2\qb\sum_{i< j}e_{12}^{[i]}e_{11}^{[j]}-2\qb\sum_{i>j}^N e_{22}^{[i]}e_{12}^{[j]}\,.
\end{equation} 

Assuming that the highest coefficients in \eqref{eq:ltexp} and \eqref{eq:cexp} are not vanishing we can take the limit $x\to\infty$ of ${\Gop}_{m,\epsilon}(x)$. We obtain 
\begin{equation}\label{eq:limG}
 \lim_{x\to\infty}{\Gop}_{m,\epsilon}(x)=\left[4 \pb \qb\left(m-e_{22}^{tot}\right) \left( \left(m-N+e_{22}^{tot}\right) - \frac{\qa}{\qb} - \frac{\pa}{\pb} \right)+\epsilon\right]^{-1} \calC_{2N-1}\,.
\end{equation} 
This yields an expression for the eigenvectors of the triangular chain without a spectral parameter. We can further simplify this expression by reordering the individual terms in the matrix powers $( \lim_{x\to\infty}{\Gop}_{m,\epsilon}(x))^k$. To do so, we note that the operator ${\calC}_{2N-1}$ satisfies the commutation relation $[e_{22}^{tot},{\calC}_{2N-1}]=-{\calC}_{2N-1}$, cf.~\eqref{eq:Ccom}. As a consequence we can commute the diagonal part in \eqref{eq:limG} through all operators ${\calC}_{2N-1}$ and act on the eigenstate of the diagonal transfer matrix $|\psi_m^0\rangle$. The  parameter $\epsilon$ can subsequently be removed.
The relation between the eigenvectors of the triangular and the diagonal spin chain can then be written as
\begin{equation}\label{eq:vecrel2}
 |\psi_m^\Delta\rangle=\sum_{k=0}^m\frac{1}{k!} \left(\frac{\Delta}{4\pb\qb}\right)^k\, \frac{ \Gamma \left(2m-N-\frac{\pa}{\pb}-\frac{\qa}{\qb}-k\right)}{\, \Gamma \left(2m-N-\frac{\pa}{\pb}-\frac{\qa}{\qb}\right)}\,\calC^k_{2N-1}|\psi^0_m\rangle\,.
\end{equation} 
Here the fraction of Gamma functions can be evaluated using $\Gamma(x+1)=x\Gamma(x)$.
This expression is much simpler than the spectral parameter dependent cousin in \eqref{eq:vecrel} as the action of the operator ${\calC}_{2N-1}$ is bi-local. Further the explicit dependence on the eigenvalue $\Lambda_m(x)$ disappeared
and only a dependence on the magnon number remains, cf.~\eqref{eq:ltexp}. This observation is equivalent to the fact that the expansion of the Baxter equation in \eqref{eq:baxter} does not contain any Bethe roots up to the order of the spectral parameter that we are interested in. 

In the next section we  evaluate \eqref{eq:vecrel2} explicitly for the case $m=N$ and derive the eigenvector of the triangular transfer matrix $|\psi_N^\Delta\rangle$ that corresponds to the other reference state $|\psi_N^0\rangle$ of the diagonal transfer matrix defined in \eqref{eq:Sigma} in closed form.

\section{Evaluation of the transformed reference state $|\psi_N^\Delta\rangle$}\label{sec:grnd}
We have seen in the previous section that the eigenvectors of triangular open spin chain can be obtained from the eigenvectors of the  diagonal open spin chain. 
While the mapping defined in \eqref{eq:vecrel} depends on the explicit form of the eigenvalue $\Lambda_m(x)$, we derived an alternative formula by taking the large $x$ limit that can be found in \eqref{eq:vecrel2}.
In this section we  study the eigenvector $|\psi_N^\Delta\rangle$ that is obtained from the reference state of the diagonal transfer matrix $|\psi_N^0\rangle$ in \eqref{eq:Sigma} which is not an eigenvector of the transfer matrix $T_\Delta(x)$ for $\Delta\neq 0$. We stress that in this case $m=N$ the eigenstate is known explicitly for any $N$ and also the corresponding eigenvalue can be computed explicitly. This is not the case for other eigenvectors of the diagonal transfer matrix. The latter are non-trivial but can be determined by e.g. the Bethe ansatz \cite{Sklyanin:1988yz}, cf.~Appendix~\ref{app:bae}. 

For the reference state $|\psi_N^0\rangle$ in \eqref{eq:Sigma}  with $m=N$ one finds the eigenvalue equation
\begin{equation}\label{eq:evt0}
 T_0(x)|\psi_N^0\rangle=\Lambda_N(x)|\psi_N^0\rangle\,,
\end{equation} 
with the transfer matrix eigenvalue explicitly given by
\begin{equation}\label{eq:ev}
 \Lambda_N(x)=\frac{2 (1 + x)^{1 + 2 N} (\pa - \pb x) (\qa - \qb x) + 
 2 x^{1 + 2 N} (\pa + \pb(x+1)) (\qa + \qb(x+1))}{1 + 2 x}\,,
\end{equation} 
see Appendix~\ref{app:fcr} for the derivation. We remark that $\Lambda_N(x)$ is by construction a polynomial in the spectral parameter $x$. The corresponding energy eigenvalue that can be obtained using the expression for the Hamiltonian \eqref{eq:hamtri} is
\begin{equation}\label{eq:enev}
 E_N^0=-\pb-\qb\,.
\end{equation}

The knowledge of the explicit form of the eigenvalue and the eigenvector in \eqref{eq:evt0} allows us to apply formula \eqref{eq:vecrel} which depends on the eigenvalue. Here, however, we focus on the eigenvalue independent relation \eqref{eq:vecrel2}. The spectral parameter independent form \eqref{eq:vecrel2} is convenient for analytic computations as the action of the appearing generators is at most bi-local. 
The evaluation of the eigenvector $|\psi_N^\Delta\rangle$ then boils down to computing the entries of the vector $|\psi_N^\Delta\rangle$ obtained from   \eqref{eq:vecrel2} for $m=N$, i.e.
\begin{equation}\label{eq:vecrel3}
 |\psi_N^\Delta\rangle=\sum_{k=0}^N\frac{1}{k!} \left(\frac{\Delta}{4\pb\qb}\right)^k\, \frac{ \Gamma \left(N-\frac{\pa}{\pb}-\frac{\qa}{\qb}-k\right)}{\, \Gamma \left(N-\frac{\pa}{\pb}-\frac{\qa}{\qb}\right)}\,\calC^k_{2N-1}|\psi_N^0\rangle\,.
\end{equation} 
As we will see the result can conveniently be written in terms of excitations on the reference state $|\psi_N^0\rangle$ of the diagonal chain.
Our final result reads
\begin{equation}\label{eq:transref}
 |\psi_N^\Delta\rangle =\sum_{h=0}^N\sum_{1\leq i_1<i_2<\ldots<i_h\leq N}\,\mathfrak{H}_\Delta(i_1,\ldots,i_h)\,
 e_{12}^{[i_1]}\cdots e_{12}^{[i_h]}|\psi_N^0\rangle\,,
\end{equation} 
with the coefficients explicitly given by
\begin{equation}\label{eq:coeffs}
 \mathfrak{H}_\Delta(i_1,\ldots,i_h)=\left(\frac{\Delta}{2\pb\qb}\right)^h\prod_{k=1}^h\frac{\qa+\qb(i_k+h-k-N)}{N-\frac{\pa}{\pb}-\frac{\qa}{\qb}-k}\,.
\end{equation} 
The derivation is presented in detail in the following two subsections.

\subsection{Coordinate form of $|\psi_N^\Delta\rangle$}\label{sec:cbalike}
In this and the next subsection we derive the explicit representation \eqref{eq:transref} and the coefficients \eqref{eq:coeffs} of the transformed reference state $|\psi_N^\Delta\rangle$.
\begin{figure}
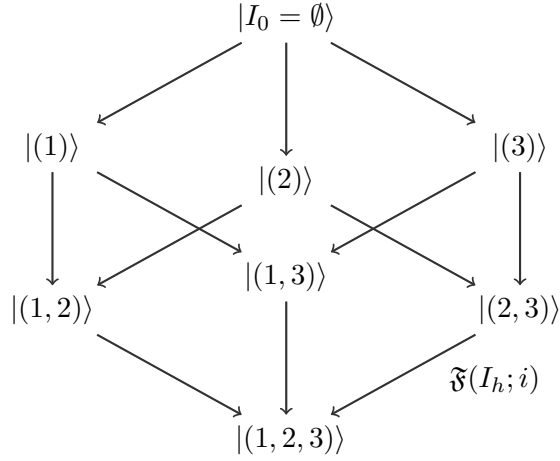

\centering
\hasse
  \caption{Hasse diagram for the case $N=3$. The filling function is associated to each arrow. For example, we get $\mathfrak{F}(2,3;1)=2(\qa+2\qb)$ for the arrow pointing from $|(2,3)\rangle$ towards $|(1,2,3)\rangle$. A path $\mathfrak{P}(i_{\sigma(1)},\ldots,i_{\sigma(h)})$ for a given permutation $\sigma\in S_h$ connects $|I_0\rangle$ to a state $|I_h\rangle$ along the direction of the arrows. The matrix element $\langle I_h |\calC^h_{2N-1}|I_0\rangle$ is obtained by summing over all such paths or equivalently permutations.}
  \label{fig:hasse}
\end{figure}

To do so, we write $|\psi_N^\Delta\rangle$ in terms of excitations on  the reference state $|\psi_N^0\rangle$. To not confuse these excitations with the ones introduced in Section~\ref{sec:base} we will call them \emph{holes}. A configuration of holes on the reference state $|\psi_N^0\rangle$ is then denoted by
\begin{equation}\label{eq:bas}
 |I_h\rangle=e_{12}^{[i_1]}\cdots e_{12}^{[i_{h}]}|\psi^0_N\rangle\,.
\end{equation} 
Here $i_k$ with $k=1,\ldots,h$ denotes the position of the $k$th hole and the ordered set $I_h$ is defined via
\begin{equation}
 I_h=(i_1,\ldots,i_h)\qquad \text{with}\qquad 1\leq i_1<\ldots<i_h\leq N.
\end{equation} 
The cardinality of $I_h$ yields the number of holes $|I_h|=h$ and for the case of $h=0$ we recover the reference state $|I_0\rangle=|\psi^0_N\rangle$.  We further note that the vectors $|I_h\rangle$ with $1\leq i_1<\ldots<i_h\leq N$ and $h=0,1,\ldots,N$ span a basis of the quantum space $V$.
To derive the form of the transformed reference state \eqref{eq:transref} we  insert the identity matrix expressed in this basis into our initial formula for $|\psi_N^\Delta\rangle$ in  \eqref{eq:vecrel3}. We find
\begin{equation}\label{eq:vecrel3b}
|\psi_N^\Delta\rangle=\sum_{h=0}^N\sum_{1\leq i_1<\ldots<i_h\leq N}\frac{\langle I_h |\calC^h_{2N-1}|I_0\rangle}{h!} \left(\frac{\Delta}{4\pb\qb}\right)^h\, \frac{ \Gamma \left(N-\frac{\pa}{\pb}-\frac{\qa}{\qb}-h\right)}{\, \Gamma \left(N-\frac{\pa}{\pb}-\frac{\qa}{\qb}\right)}\,|I_h\rangle\,.
\end{equation} 
It remains to evaluate the matrix elements $\langle I_h |\calC^h_{2N-1}|I_0\rangle$.
As presented below, this can be done by deriving the action of the operator ${\calC}_{2N-1}$ on the $|I_h\rangle$. 

Using the explicit expression for ${\calC}_{2N-1}$ in \eqref{eq:cexp} we find that ${\calC}_{2N-1}$ acting on $|I_h\rangle$ yields a sum of terms with an additional hole at every possible position $i$, denoted by $|I_h\cup (i)\rangle$, as long as the position $i$ is not yet already occupied by a hole $i\notin I_h$. The operator  ${\calC}_{2N-1}$  thus raises the number of holes by one and its action on $|I_h\rangle$ can be written as
\begin{equation}\label{eq:mapC}
{\calC}_{2N-1}|I_h\rangle=\sum_{i=1}^N\left[1-w_i(I_h)\right] \mathfrak{F}(I_{h};i)|I_h\cup (i)\rangle\,.
\end{equation} 
Here we defined the function $w_i(I_h)$ that returns  $w_i(I_h)=1$ if there is a hole at position $i$ and $w_i(I_h)=0$ if not. It can be written as the sum
\begin{equation}
 w_j(I_h)=\sum_{k=1}^h \delta_{i_k,j}\,.
\end{equation} 
A direct computation shows that the coefficients of the basis vectors $|I_h\cup (i)\rangle$ in \eqref{eq:mapC} are given in terms of the filling function
\begin{equation}
 \mathfrak{F}(I_{h};i)=2\left(\qa+\qb\sum_{j=i+1}^N(2w_j(I_{h})-1)\right)\,.
\end{equation} 
It is convenient to visualise the action of ${\calC}_{2N-1}$ in a Hasse diagram, see Figure~\ref{fig:hasse} for the case with $N=3$ spin chain sites. The nodes of the diagram represent the basis vectors $|I_h\rangle$ with $h=0,1,2,3$. Here the basis vectors $|I_h\rangle$ on each level of the diagram have a fixed number of holes. The operator ${\calC}_{2N-1}$ then lowers the level. It maps a given state $|I_h\rangle$ to an overlap of all states that are pointed at by the arrows and originate from $|I_h\rangle$. To each arrow from  $|I_{h}\rangle$ to $|I_{h}\cup (i)\rangle$ we associate the filling function $\mathfrak{F}(I_{h},i)$.

In order to compute the action ${\calC}_{2N-1}^h|I_0\rangle$ we define the function 
\begin{equation}
 \mathfrak{P}(i_{\sigma(1)},\ldots,i_{\sigma(h)})=\prod_{k=0}^{h-1}\mathfrak{F}(i_{\sigma(1)},\ldots,i_{\sigma(k)};i_{\sigma(k+1)})\,,
\end{equation} 
which corresponds to a path from $|I_0\rangle$ to $|I_h\rangle$ on the Hasse diagram. For a given state $|I_h\rangle$ there are $h!$ different paths originating from $|I_0\rangle$. They are labelled by a permutation $\sigma\in S_h$. The coefficient of ${\calC}_{2N-1}^h|I_0\rangle$ that is proportional to $|I_h\rangle$ is then obtained by summing over all possible paths or equivalently over all permutations. We obtain
\begin{equation}\label{eq:permsum}
\langle I_h|{\calC}^h_{2N-1}|I_0\rangle=\sum_{\sigma\in S_h} \mathfrak{P}(i_{\sigma(1)},\ldots,i_{\sigma(h)})\,.
\end{equation} 
Remarkably we can further simplify \eqref{eq:permsum} by summing up the permutations. We obtain
\begin{equation}\label{eq:noperm}
\begin{split}
\langle I_h|{\calC}^h_{2N-1}|I_0\rangle
&=2^h h!\prod_{\alpha=1}^h(\qa+\qb(i_\alpha+h-\alpha-N))\,,
\end{split}
\end{equation} 
and immediately arrive at \eqref{eq:transref}.
This last step is shown in the next subsection.

\subsection{Summing over permutations}\label{sec:permut}
In the following subsection we show that the sum over the permutations in  \eqref{eq:permsum} can be simplified to  the simple product spelled out in \eqref{eq:noperm}.

By construction the coefficients $\langle I_h|\calC^h_{2N-1}|I_0\rangle$ written in terms of the sum over the permutations in \eqref{eq:permsum}
satisfy  the recursion relation 
\begin{equation}\label{eq:rec}
\langle I_h|\calC^h_{2N-1}|I_0\rangle=\sum_{k=1}^h\mathfrak{F}(I_h\setminus (i_k);i_k)\langle I_h\setminus(i_k)|\calC^{h-1}_{2N-1}|I_0\rangle\,,
\end{equation} 
which relates two neighboring levels of the Hasse diagram. Here $I_h\setminus (i_k)$ denotes the ordered set $I_h\setminus (i_k)=(i_1,\ldots,i_{k-1},i_{k+1},\ldots,i_h)$. The sum runs over all possibilities to build the state $|I_h\rangle$ from the level with $h-1$ holes. 

To show that \eqref{eq:permsum} coincides with \eqref{eq:noperm}
we first validate that this is the case for $h=0$. In both cases we find 
\begin{equation}
 \langle I_0|\calC^0_{2N-1}|I_0\rangle=1\,.
\end{equation} 
It then remains to verify that the recursion relation \eqref{eq:rec} is satisfied by the coefficients \eqref{eq:noperm}. To do so we first remark that we can evaluate the sum in  the filling function 
\begin{equation}
 \mathfrak{F}\left(I_h\setminus (i_k);i_k\right)=2\left(\qa+\qb(2(h-k)+i_k-N)\right)\,,
\end{equation} 
keeping in mind that $i_1<\ldots< i_h$.  When substituting this form of $\mathfrak{F}\left(I_h\setminus (i_k);i_k\right)$ into \eqref{eq:rec} along with with the coefficients given in \eqref{eq:noperm} we confirm that
\begin{equation}
\begin{split}
\sum_{k=1}^h\mathfrak{F}\left(I_h\setminus (i_k);i_k\right)&\langle I_h\setminus (i_k)|\calC^{h-1}_{2N-1}|I_0\rangle =2^{h}(h-1)!\sum_{k=1}^h\left(\qa+\qb(2(h-k)+i_k-N)\right)\\&\quad\times\Bigg[\prod_{j=1}^{k-1}(\qa+\qb(i_j+h-1-j-N))
\prod_{j=k+1}^{h}(\qa+\qb(i_j+h-j-N))\Bigg]\\&=2^h h!\prod_{j=1}^{h}(\qa+\qb(i_j+h-j-N))\\&=\langle I_h|\calC^{h}_{2N-1}|I_0\rangle\,.
\end{split}
\end{equation} 
The relation above relies on the identity
\begin{equation}
\begin{split}
\sum_{k=1}^h\frac{\qa+\qb(2(h-k)+z_k-N)}{\qa+\qb(z_k+h-k-N)}\prod_{j=1}^{k-1}\frac{\qa+\qb(z_j+h-1-j-N)}{\qa+\qb(z_j+h-j-N)}=h\,,
\end{split}
\end{equation} 
which holds for any $z_k\in\mathbb{C}$.
It can be shown by taking
\begin{equation}
 z_k =k-h+ N - \frac{\qa}{\qb} + \frac{1}{1 - \tilde z_k}\,.
\end{equation} 
One  finds
\begin{equation}
\begin{split}
\sum_{k=1}^h\frac{\qa+\qb(2(h-k)+z_k-N)}{\qa+\qb(z_k+h-k-N)}&\prod_{j=1}^{k-1}\frac{\qa+\qb(z_j+h-1-j-N)}{\qa+\qb(z_j+h-j-N)}\\&=
\sum_{k=1}^h(1 + h - k +(k- h) \tilde z_k)\prod_{j=1}^{k-1}\tilde z_j
\\&=h+\sum_{k=2}^h(1 + h - k)\prod_{j=1}^{k-1}\tilde z_j+\sum_{k=1}^{h-1}(k- h) \prod_{j=1}^{k}\tilde z_j
\\&=h\,.
\end{split}
\end{equation} 
This concludes the proof of the eigenvector formula \eqref{eq:transref} with the coefficients \eqref{eq:coeffs} for $|\psi_N^\Delta\rangle$.

\section{Some implications for the open SSEP}\label{sec:ssep}
In this section we apply our results to the Symmetric Simple Exclusion Process with reservoirs (open SSEP). For this purpose we first review the relation among the triangular spin chain defined in Section~\ref{sec:base} and the open SSEP in Section~\ref{sec:ssepqism}. We closely follow the reference \cite{Crampe:2014aoa}. Our results are then presented in Section~\ref{sec:ssepresuts}. Here we give a new representation of the eigenvectors and in particular the steady state \eqref{eq:steadystate}, the components of the steady state \eqref{eq:probs} and the steady state correlation functions \eqref{eq:kptfct} of the open SSEP using the results derived in the previous sections for the triangular chain.

\subsection{The open SSEP in the framework of the QISM}\label{sec:ssepqism}

The symmetric simple exclusion process with reservoirs is generated by the Markov matrix
\begin{equation}\label{eq:ssepham}
\tilde H=\tilde H_1^L+\sum_{k=1}^{N-1}(P_{k,k+1}-\ID)+\tilde H_N^R\,,
\end{equation} 
where the permutation operator can be expressed in terms of the Pauli matrices via $P_{k,k+1}=\frac{1}{2}(\vec\sigma^{[k]}\cdot \vec\sigma^{[k+1]} +\ID)  $ and the boundary terms read
\begin{equation}\label{eq:bndham}
\tilde H^L=\left(\begin{array}{cc}
                   -\alpha&\gamma\\
                   \alpha&-\gamma
                  \end{array}
\right)
 \,,\qquad  \tilde H^R=\left(\begin{array}{cc}
                   -\delta&\beta\\
                   \delta&-\beta
                  \end{array}
\right)\,.
\end{equation} 
The Markov matrix in \eqref{eq:ssepham} can be identified with the Hamiltonian of an open Heisenberg chain and thus we will also refer to it as the Hamiltonian of the open SSEP.
It describes a process where  particles may hop to the left and right in the bulk with rate ``$1$'', while particles are inserted at the left (right) of the chain at rate $\alpha$ ($\delta$) and removed at rate $\gamma$ ($\beta$). These particles can be identified with the magnons introduced in Section~\ref{sec:base}.

The open SSEP can be formulated within the framework of the quantum inverse scattering method. The underlying transfer matrix is constructed as explained in Section~\ref{sec:base} using different boundary K-matrices
\begin{equation}
 \tilde T(x)=
 \tr_a \tilde K_a(x)\tilde U_a(x)\quad\text{with}\quad\tilde U_a(x)=M_a(x)\tilde{\hat K}_a(x)\hat M_a(x)\,.
\end{equation} 
The single row monodromies were defined in \eqref{eq:singmon}. The K-matrices are naively neither  triangular nor diagonal. They are of the form
\begin{equation}\label{eq:kmat}
 \tilde K(x)=\left(
\begin{array}{cc}
 1+(x+1)(\gamma -\alpha )& 2 (x+1) \gamma  \\
 2 (x+1) \alpha  &1+ (x+1) (\alpha -\gamma ) \\
\end{array}
\right),\;\;\tilde{\hat{ K}}(x)=\left(
\begin{array}{cc}
 1+x (\beta -\delta )  & 2 x \beta  \\
 2 x \delta & 1+x (\delta -\beta )\\
\end{array}
\right)
\end{equation} 
and satisfy the boundary Yang-Baxter equation. The transfer matrix constructed from these K-matrices is related to the Hamiltonian in \eqref{eq:ssepham} via
\begin{equation}\label{eq:logder}
 \tilde H =\frac{1}{2}\left(\frac{\partial}{\partial x}\log \tilde T(x)|_{x=0}-\left(2N-1+\alpha+\beta+\delta+\gamma\right)\ID\right)\,.
\end{equation} 
It follows that the transfer matrix commutes with the Hamiltonian and that they share the same set of eigenvectors. 

In general it is rather involved to obtain the eigenvectors of such spin chains where the K-matrices do not have any non-vanishing entries. However for the open SSEP it is known that the problem can be mapped to the triangular spin chain, see \cite{PhysRevLett.95.240601,degier2006,Melo:2005gut,Crampe:2014aoa}.
More precisely introducing the transformation matrix 
\begin{equation}\label{eq:Gamma}
 \Gamma=\left(
\begin{array}{cc}
 -\frac{1}{\alpha+\gamma} & \frac{\beta}{\beta +\delta } \\
 \frac{1}{\alpha+\gamma} & \frac{ \delta }{\beta +\delta } \\
\end{array}
\right)\,,
\end{equation} 
one finds that $T_\Delta$ is similar to $\tilde T$, i.e.
\begin{equation}\label{eq:ttt}
 T_\Delta(x)=\mathcal{S}_\Gamma^{-1}\tilde T(x)\mathcal{S}_\Gamma\,,
\end{equation} 
where $\mathcal{S}_\Gamma=\Gamma\otimes\Gamma\otimes\ldots\otimes \Gamma$ and we identified the parameters of the triangular chain with the ones of the open SSEP via
\begin{equation}\label{eq:idenf}
p=-(\alpha+\gamma)\,,\qquad \Delta=2\frac{(\alpha +\gamma)(\alpha  \beta -\gamma  \delta )}{\beta +\delta  } \,,\qquad q=-(\beta+\delta)\,.
\end{equation} 
We assume that this indentification of the parameters holds throughout the remaining part of this section.
The identification for the transfer matrices in \eqref{eq:ttt} follows noting that the K-matrices in \eqref{eq:kmat} satisfy
\begin{equation}
\Gamma^{-1}\tilde K(x)\Gamma=\left(
\begin{array}{cc}
 1-(x+1) (\alpha + \gamma)  &\frac{2 (x+1) (\alpha +\gamma ) (\alpha  \beta -\gamma  \delta )}{\beta +\delta } \\
 0 &1+(x+1)( \alpha +\gamma) \\
\end{array}
\right)\,,
\end{equation} 
and 
\begin{equation}
 \Gamma^{-1}\tilde {\hat K}(x)\Gamma=\left(
\begin{array}{cc}
 1-x (\beta +\delta ) & 0 \\
 0 & 1+x (\beta +\delta ) \\
\end{array}
\right)\,,
\end{equation} 
as well as the invariance relation  $[R(x),\Gamma\otimes\Gamma]=0$ holds.

As a consequence the Hamiltonian of the open SSEP \eqref{eq:ssepham} can be brought to the triangular form given in \eqref{eq:triham} under the identification of the parameters in \eqref{eq:idenf}.  The transformation \eqref{eq:univham2} is defined through 
\begin{equation}
 c_1=2\,,\qquad c_0=\alpha+\beta+\gamma+\delta\,,\qquad G=\Gamma\,,
\end{equation} 
with the matrix $\Gamma$ defined in \eqref{eq:Gamma}.

In the next subsections we  define the eigenstates in terms of the transformed reference state $|\psi_N^\Delta\rangle$, discuss the  representation of the steady state, its components and the steady state correlation functions that arises from it.

\subsection{Eigenvectors, steady state and correlation functions}\label{sec:ssepresuts}

Given the relation of the transfer matrix of the open SSEP and the transfer matrix of the triangular chain in \eqref{eq:ttt} we can express the eigenvectors of the open SSEP in terms of the eigenvectors of the triangular chain (which we had previously expressed in terms of the eigenvectors of the diagonal chain). We have
\begin{equation}
 \tilde T(x)|\tilde\psi_m\rangle =\tilde\Lambda_m(x)|\tilde\psi_m\rangle\,,
\end{equation} 
with $\tilde \Lambda_m(x)=\Lambda_m(x)$ and 
\begin{equation}\label{eq:steadystate}
 |\tilde\psi_m\rangle =\mathcal{S}_\Gamma|\psi_m^\Delta\rangle\,.
\end{equation} 
Here we assume the identification of the parameters as given in \eqref{eq:idenf}.

Of particular interest is the steady state which is a null vector of the open SSEP Hamiltonian in \eqref{eq:ssepham}. This should be the case for the state with $m=N$, see e.g.~\cite{Crampe:2014aoa}.
In fact, this is straightforward to verify. The eigenvalue of the Hamiltonian follows from \eqref{eq:logder} together with the energy eigenvalue of the triangular chain spelled out in \eqref{eq:enev}. Under the identification of the parameters in \eqref{eq:idenf} one finds
\begin{equation}
 \tilde H|\tilde \psi_N\rangle =0\,,
\end{equation} 
which is the defining relation for the steady state.

As mentioned above, the magnons in Section~\ref{sec:base} are now interpreted as particles such that the occupation number at site $i$ is given by the local magnon number $m_i$   at site $i$.
The probabilities to find a configuration $m_1,\ldots,m_N$ in the steady state are simply given by the components of the eigenvector
\begin{equation}\label{eq:prob}
 \PP(m_1,\ldots,m_N)=\langle m_1,\ldots,m_N|\tilde\psi_N\rangle\,.
\end{equation} 
Here $\langle m_1,\ldots,m_N|$ denotes the state with magnon number $m_i$ at the $i$th site, cf. Section~\ref{sec:base}. In particular we have $\langle 0,\ldots,0|=|\psi_0^\Delta\rangle^t$.
This matches with \cite{derrida2007non} where ``$1$'' denotes a particle and ``$0$'' an empty site. The probabilities can then be computed case by case using the DEHP matrix product ansatz as summarised in Appendix~\ref{sec:dehp} or in principle, as discussed in the introduction, by Bethe ansatz.

In the following we take a different route to obtain the probabilities \eqref{eq:prob} and derive an analytic expression for it. First we validate that the probabilities are properly normalised, i.e.
\begin{equation}
 \sum_{\{m_i\}=0,1}\PP(m_1,\ldots,m_N)=\langle 1,\ldots,1|\psi_N^\Delta\rangle=1\,.
\end{equation} 
This can be shown using the identification in \eqref{eq:idenf} and the relation
\begin{equation}\label{eq:Gproj}
 \left(\langle 0|+\langle 1|\right)\Gamma=\langle 1|\,.
\end{equation} 
Next we insert the explicit expression for the transformed reference state in \eqref{eq:transref} with the coefficients in \eqref{eq:coeffs} into the definition of the probabilities \eqref{eq:prob}. This representation allows us to derive an exact form of the probabilities for arbitrary $N$. Noting that the action of the matrix $\Gamma$ in \eqref{eq:Gamma} on the basis vectors can be written as
\begin{equation}
 \langle m|\Gamma=\frac{(-1)^{m+1}}{\alpha+\gamma}\langle 0|+\left(\frac{\delta}{\beta}\right)^m\frac{\beta}{\beta+\delta}\langle 1|\,,\qquad\text{for}\qquad m=0,1\,,
\end{equation} 
we obtain
\begin{equation}
 \langle m_1,\ldots,m_N|\mathcal{S}_\Gamma=\frac{\delta^m\beta^{N-m}}{(\beta+\delta)^N}\sum_{h=0}^N\left(-\frac{\beta+\delta}{\beta(\alpha+\gamma)}\right)^h\sum_{1\leq i_1<\ldots<i_h\leq N}\left(-\frac{\beta}{\delta}\right)^{\sum_{k=1}^h m_{i_k}}\langle i_1,\ldots,i_h|\,,
\end{equation} 
with $m=\sum_{i=1}^N m_i$ and $\langle i_1,\ldots,i_h|$ defined as the transposed vectors in \eqref{eq:bas}. 
This immediately yields the expression for the probabilities.
 After using the identification in \eqref{eq:idenf} we find
\begin{equation}\label{eq:probs}
 \PP(m_1,\ldots,m_N)=\sum_{h=0}^N(1-\rho_b)^{N-h}\left(\rho_a-\rho_b\right)^h\sum_{ 1\leq i_1<\ldots<i_h\leq N}\mathfrak{M}(i_1\ldots,i_h;m_1,\ldots,m_N)\,,
\end{equation} 
with the coefficients simply given by
\begin{equation}
\mathfrak{M}(i_1\ldots,i_h;m_1,\ldots,m_N)=(-1)^{\sum_{k=1}^h m_{i_k}}
\left(\frac{\delta}{\beta}\right)^{m-\sum_{k=1}^h m_{i_k}}
 \prod_{k=1}^h\frac{i_{k}+h-k-N-\frac{1}{\beta+\delta}}{N-k+\frac{1}{\alpha+\gamma}+\frac{1}{\beta+\delta}}\,.
\end{equation} 
Further we have introduced the densities of the left and right reservoirs
\begin{equation}
 \rho_a=\frac{\alpha}{\alpha+\gamma}\,,\qquad\rho_b=\frac{\delta}{\beta+\delta}\,.
\end{equation} 
In particular for $\rho_a=\rho_b=\rho$ the only term that survives in the sum of for the probabilities in \eqref{eq:probs} is the one with $h=0$. In this case the system is in equilibrium and we recover the Bernoulli measure at density $\rho$, i.e.
$
 \PP(m_1,\ldots,m_N)=(1-\rho)^{N-m}\rho^m\,,
 $
cf.~\cite{derrida2007non}.

We can also obtain an analytic expression  for the steady state correlation functions. These can be defined as 
\begin{equation}\label{eq:cordef}
\begin{split}
 \langle{i_1}\cdots {i_k}\rangle&=\sum_{\{m_j|j\neq i_1,\ldots, i_k\}=0,1} \PP(m_1,\ldots,m_N)\bigg|_{m_{i_1}=\ldots=m_{i_k}=1}\,.
 \end{split}
\end{equation} 
Here we fixed the particle numbers $m_{i_1}=\ldots =m_{i_k}=1$ and sum over the different configurations at the remaining sites. This sum in \eqref{eq:cordef} can be written in terms of the vector $|\psi_N^\Delta\rangle$ in \eqref{eq:transref} with the identifications in \eqref{eq:idenf} as
\begin{equation}
 \langle{i_1}\cdots {i_k}\rangle=\langle I_0|\Gamma^{[i_1]}\cdots\Gamma^{[i_k]}|\psi^\Delta_N\rangle\,,
\end{equation} 
with $\langle I_0|=\langle 1,\ldots,1|$ and where we have used the projection property of the matrix $\Gamma$ given in \eqref{eq:Gproj}.
Noting  that the action of the remaining matrices $\Gamma$ can be written as
\begin{equation}
 \langle I_0|\Gamma^{[i_1]}\cdots\Gamma^{[i_k]}=\sum_{j=0}^k\frac{1}{(\alpha+\gamma)^{j}}\frac{\delta^{k-j}}{(\beta+\delta)^{k-j}}\sum_{1\leq w_1<\ldots< w_j\leq k}\langle i_{w_1},\ldots ,i_{w_j}|\,,
\end{equation} 
we obtain an explicit formula for the $k$-point correlation function with arbitrary $N$. We find
\begin{equation}\label{eq:kptfct}
\begin{split}
 \langle{i_1}\cdots{i_k}\rangle
 &=\sum_{m=0}^k(\rho_b-\rho_a)^m(\rho_b)^{k-m}\sum_{1\leq l_1<\ldots<l_m\leq k}\,\prod_{r=1}^m\frac{i_{l_r}+m-r-N-\frac{1}{\beta+\delta}}{N-r+\frac{1}{\alpha+\gamma}+\frac{1}{\beta+\delta}}\,.
 \end{split}
\end{equation} 
In particular, for $k=1$, this gives the well known expression for the steady state density profile
\begin{equation}
 \langle i\rangle=\frac{\rho_a\left(N+\frac{1}{\beta+\delta}-i\right)+\rho_b\left(i-1+\frac{1}{\alpha+\gamma}\right)}{N+\frac{1}{\alpha+\gamma}+\frac{1}{\beta+\delta}-1}\,.
\end{equation} 
This result has been obtained using the DEHP ansatz in \cite{Derrida_2002}. Higher point correlations were given in \cite{Derrida_2006} where also a recursion relation which relates $k$ points to $k-1$ points can be found. However, to our knowledge, closed-form expressions as given in this section for the particle distributions  \eqref{eq:probs}  and  correlations \eqref{eq:kptfct}  in the steady state have not appeared previously in the literature.

\section{Conclusions}\label{sec:conc}
In this article we have presented formulas for the eigenvectors of the open Heisenberg spin chain with triangular boundary conditions in terms of the eigenvectors of the open spin chain with diagonal boundary conditions. The relation with and without the spectral parameter are given in \eqref{eq:vecrel} and \eqref{eq:vecrel2}.  In general for the eigenstates beyond half-filling our method allows to use the symmetries of the diagonal transfer matrix where eigenstates can be obtained from the Bethe ansatz on two different reference states. Further we got an explicit expression for the transformed reference state $|\psi_N^\Delta\rangle$ in \eqref{eq:transref} and \eqref{eq:coeffs} which we obtained by acting on the reference state of the diagonal chain $|\psi^0_N\rangle$ with the generators of the Yangian and without solving any Bethe equations.

We then argued that the eigenstates of the open SSEP can be obtained from the eigenstates of the triangular and thus from the eigenstates of the diagonal open spin chain. The transformed reference state $|\psi^\Delta_N\rangle$  was shown to yield the steady state for which we deduced the probabilities for a given configuration in a closed form \eqref{eq:probs} and also derived an exact expression for the correlation functions \eqref{eq:kptfct}. Remarkably, to our knowledge, such concise expressions were not obtained previously in the literature.

It would be interesting to study whether similar results can be obtained for the XXZ chain and the open ASEP whose Hamiltonian and Markov generator are known to be related by a similarity transformation, see e.g.~\cite{schutz2000exactly}. The algebraic Bethe ansatz for triangular spin chains of this type have been studied in \cite{Pimenta:2013cua,Belliard:2014fsa} and the DEHP ansatz works for ASEP without any major changes. One would expect that the algebraic expressions for the eigenvectors in Section~\ref{sec:evs} do immediately carry over to the trigonometric spin chain. The main difference compared to the case studied here is that the R-matrix is no longer $\mathfrak{gl}(2)$ invariant which we have used to bring the transfer matrix of the open SSEP to a triangular form. 

We expect that the method straightforwardly generalises to higher spin Heisenberg chains. In particular it would be very interesting to study the non-compact case which was recently shown to be related to a stochastic process without exclusion in \cite{Frassek2019}, see also \cite{Frassek:2019isa} for the asymmetric version for which however no integrable stochastic boundary conditions are known.
So far we have only investigated some of the microscopic properties for the SSEP that arise from the new representation of the transformed reference state. It would be interesting to study the macroscopic limit and the role of duality.

We further remark that in the framework of the QISM the probabilities of the configurations in the steady state can be interpreted as some sort of scalar product. The latter have been studied extensively and  one may expect that one may be able to obtain yet another representation of the components of the  steady state.
It could be interesting to employ the method of separation of variables, see e.g. \cite{Kitanine:2016pvg} as well as \cite{Maillet:2019hdq} and references therein.

Another promising direction is to apply the method presented in the article to other models that can be solved by a  DEHP matrix product ansatz. In particular it is tempting to investigate whether one can obtain explicit formulas for the multi-species versions of the SSEP and ASEP as well as TASEP, see e.g.~\cite{Crampe,vanicat2017exact,finn2018matrix}, using the relation to higher rank spin chains for which the transformed reference state may also be computed exactly. 

Finally we hope that the eigenstates of general open spin chains can in some similar way be related to the triangular case and thus the diagonal case studied here. It is however known that the this case is rather different to the one studied here. The eigenvectors can be obtained through the modified Bethe ansatz in \cite{Belliard:2013aaa,Cao:2014aqa} and the spectrum is known to be described by an inhomogeneous TQ-equation, see in particular \cite{Cao:2013qxa,Nepomechie:2013ila}.

\paragraph{Acknowledgements}
I like to thank Bernard Derrida, Eric Ragoucy,  Jens Niklas Eberhardt, Jorge Kurchan, Rafael Nepomechie and especially Cristian Giardin\`{a} and Rodrigo A. Pimenta for very interesting discussions and helpful comments. Further I thank the anonymous referees for their useful comments on the manuscript. Finally, I thank the Mathematical and Statistical Physics group of Bergische Universit\"at Wuppertal, where parts of this work were presented in a seminar, for comments and acknowledge the support of the IH\'ES visitor program. 
I also acknowledge the support of the DFG Research Fellowships Programme 416527151.

\appendix

\section{Eigenvalue of reference state}\label{app:fcr}
In this appendix we derive the eigenvalue $\Lambda_N(x)$ of the transfer matrix $T_0(x)$ for $m=N$ as presented in \eqref{eq:ev}. 
First we note that the single row monodromies $M,\hat M$ satisfy the Yang-Baxter relation
\begin{equation}\label{eq:rtt2}
 \hat M_a(x)R_{ab}(x+y)M_b(y)=M_b(y)R_{ab}(x+y)\hat M_a(x)\,.
\end{equation} 
This is a consequence of the ordinary RTT-relation
\begin{equation}\label{eq:rtt1}
 R_{ab}(x-y)M_a(x)M_b(y)=M_b(y)M_a(x)R_{ab}(x-y)\,.
\end{equation} 
More precisely, we obtain \eqref{eq:rtt2} when transposing the RTT-relation \eqref{eq:rtt1} in the first space ``$a$'' and using that
\begin{equation}
\begin{split}
 M^{t_a}_a(x)=(-1)^{N-1}\left(\begin{array}{cc}
             0&-1\\
             1&0
            \end{array}
\right)\hat M_a(-x-1)\left(\begin{array}{cc}
             0&-1\\
             1&0
            \end{array}
\right)\,,
\end{split}
\end{equation}
which relies on the crossing relation of the R-matrix 
\begin{equation}
 R_{a,b}(z)=\left(\begin{array}{cc}
             0&1\\
             -1&0
            \end{array}
\right)R_{a,b}^{t_a}(-z-1)\left(\begin{array}{cc}
             0&1\\
             -1&0
            \end{array}
\right)\,.
\end{equation} 
Now we define
\begin{equation}
 M(x)=\left(\begin{array}{cc}
             A(x)&B(x)\\
             C(x)&D(x)
            \end{array}
\right)\,,\qquad  \hat M(x)=\left(\begin{array}{cc}
             \hat A(x)&\hat B(x)\\
             \hat C(x)&\hat D(x)
            \end{array}
\right)\,.
\end{equation} 
The diagonal entries of the double-row monodromy \eqref{eq:drow} can then be written as
\begin{equation}
 \calA(x)=(\qa+x\qb)A(x)\hat A(x)+(\qa-x\qb)B(x)\hat C(x)\,,
\end{equation} 
\begin{equation}
 \calD(x)=(\qa-x\qb)D(x)\hat D(x)+(\qa+x\qb)C(x)\hat B(x)\,.
\end{equation} 
Further from \eqref{eq:rtt2} we obtain the commutation relations
\begin{equation}
 B(x)\hat C(x)=\hat C(x)B(x)+\frac{1}{2x+1}\left(\hat D(x) D(x)-A(x)\hat A(x)\right)\,.
\end{equation} 
We can now compute the action of the of $\mathcal{A}(x)$ and $\mathcal{D}(x)$ on the reference state. Noting that 
\begin{equation}
 A(x)|\psi^0_N\rangle = \hat A(x)|\psi^0_N\rangle =x^N|\psi^0_N\rangle \,,
\end{equation} 
\begin{equation}
 D(x)|\psi^0_N\rangle = \hat D(x)|\psi^0_N\rangle =\left(x+1\right)^{N}|\psi^0_N\rangle \,,
\end{equation} 
\begin{equation}
 B(x)|\psi^0_N\rangle=\hat B(x)|\psi^0_N\rangle =0\,,
\end{equation} 
we find 
\begin{equation}
 \calA(x)|\psi^0_N\rangle =(\qa+x\qb)x^{2N}+\frac{\qa-x\qb}{2x+1}\left(\left(x+1\right)^{2N}-x^{2N}\right)\,,
\end{equation} 
and
\begin{equation}
 \calD(x)|\psi^0_N\rangle =(\qa-x\qb)\left(x+1\right)^{2N}\,.
\end{equation} 
This yields the eigenvalue $\Lambda_N(x)$ in \eqref{eq:ev}.
\section{Bethe ansatz for the diagonal spin chain}\label{app:bae}
The eigenvectors of the diagonal transfer matrix $T_0(x)$ can be obtained using one of the reference state $|\psi_{0}^{0}\rangle=|\psi_{0}^{\Delta}\rangle$ as defined in \eqref{eq:ref} or $|\psi_N^0\rangle$ as given in \eqref{eq:Sigma} by using the standard algebraic Bethe ansatz \cite{Sklyanin:1988yz}. They are given up to a normalisation in terms of the off-diagonal entries of the double-row-monodromy \eqref{eq:drowabcd} via
\begin{equation}\label{eq:cops}
 |\psi_{m_+}^0\rangle\propto\mathcal{C}(x_1^+)\cdots \mathcal{C}(x_{m_+}^+)|\psi_N^0\rangle\,,
\end{equation}
or alternatively
\begin{equation}\label{eq:bops}
 |\psi_{m_-}^0\rangle\propto\mathcal{B}(x_1^-)\cdots \mathcal{B}(x_{m_-}^-)|\psi_0^0\rangle\,.
\end{equation} 
Here the magnon numbers are denoted as $m_\pm=0,1,\ldots,N$ and denote the number of  excitations with respect to the  chosen reference state. The Bethe roots $x_i^\pm$ with $i=1,\ldots,m_\pm$ are solutions of the Bethe equations
\begin{equation}\label{eq:bae}
\frac{\left(\pa\pm x_i^\pm\pb\right)\left(\qa\pm x_i^\pm \qb\right)\left(x_i^\pm+1\right)^{2N}}{\left(\pa\mp\left(x_i^\pm+1\right)\pb\right)\left(\qa\mp\left(x_i^\pm+1\right)\qb\right)\left(x_i^\pm\right)^{2N}}=\prod_{\substack{k=1 \\ k\neq i}}^{m_\pm}\frac{(x_i^\pm-x_k^\pm+1)(x_i^\pm+x_k^\pm+2)}{(x_i^\pm-x_k^\pm-1)(x_i^\pm+x_k^\pm)}\,.
\end{equation} 
 The corresponding eigenvalues of the transfer matrix $T_0(x)$ can then be written in terms of the Baxter equation as
\begin{equation}\label{eq:baxter}
\begin{split} 
 \Lambda(x)&=(\pa\pm x\pb)(\qa\pm x \qb)\frac{2}{2x+1}\left(x+1\right)^{2N+1}\frac{Q^{\pm}(x-1)}{Q^{\pm}(x)}\\
 &\quad+(\pa\mp (x+1)\pb)(\qa\mp(x+1)\qb)\frac{2}{2x+1}x^{2N+1}\frac{Q^\pm(x+1)}{Q^\pm(x)}\,,
 \end{split}
\end{equation} 
with the Q-functions  defined as
\begin{equation}
 Q^{\pm}(x)=\prod_{i=1}^{m_\pm} (x-x_i^\pm)(x+x_i^\pm+1)\,.
\end{equation} 
To match the  conventions used for the magnon excitations in the main text we have to identify $m=m_-=N-m_+$.  

Let us further comment on difference of choosing the representations of the eigevectors   \eqref{eq:cops} or \eqref{eq:bops} in the determination of the transformed reference state $|\psi^\Delta_N\rangle$. The transformed reference state is obtained from $|\psi^0_N\rangle$ via \eqref{eq:vecrel3}. The state $|\psi^0_N\rangle$ can either be obtained from \eqref{eq:cops} with $m_+=0$ or from \eqref{eq:bops} with $m_-=N$. The first case does not require to solve any Bethe equations and is thus the natural choice. In fact, it seems advantageous to choose the representation \eqref{eq:cops} to \eqref{eq:bops} as long as $m\geq\lceil\frac{N+1}{2}\rceil$. This flexibility of choosing \eqref{eq:cops} or \eqref{eq:bops} is absent in the modified Bethe ansatz approach \cite{BCR}.   When fixing the reference state as in \eqref{eq:bops}, we should recover the Bethe off-shell vector as given in \cite{BCR} after homogeneously distributing the transformation defined as part of \eqref{eq:vecrel} among the operators $\mathcal{B}(x_i^-)$. It would be interesting to verify this relation explicitly for arbitrary magnon blocks.

\section{DEHP matrix product ansatz}\label{sec:dehp}
Following \cite{Derrida:1992vu} we define
\begin{equation}\label{eq:dehppro}
 \PP(m_1,\ldots,m_N)=\frac{\langle W|X_1\cdots X_N|V\rangle}{\langle W|(E+D)^N|V\rangle}\,,
\end{equation} 
where 
\begin{equation}
 X_i=m_iD+(1-m_i)E\,.
\end{equation} 
Here $D$ denotes an occupied site and $E$ an empty site. The operators $D$ and $E$ satisfy the commutation relations
\begin{equation}
 DE-ED=D+E\,.
\end{equation}
Further at the boundary one imposes the conditions
\begin{equation}
 \langle W|(\alpha E-\gamma D)=\langle W|\,,\qquad (\beta D-\delta E)| V\rangle=|V\rangle\,.
\end{equation} 
It is convenient to pick an explicit realisation to compute the probabilities in \eqref{eq:dehppro}. We use the one proposed in \cite{wadati}. It is given by the infinite-dimensional matrices 
\begin{equation}
 D=\left(\begin{array}{cccc}
            D_{11}&D_{12}&0&\cdots\\
            D_{21}&D_{22}&D_{23}&\\
            0&D_{32}&D_{33}&\ddots\\
            \vdots&&\ddots&\ddots
           \end{array}
\right)\,,\qquad  E=\left(\begin{array}{cccc}
            E_{11}&E_{12}&0&\cdots\\
            E_{21}&E_{22}&E_{23}&\\
            0&E_{32}&E_{33}&\ddots\\
            \vdots&&\ddots&\ddots
           \end{array}
\right)         \,,
\end{equation} 
along with the infinite-dimensional vectors
\begin{equation}
\langle W|=(1,0,0,\ldots)\,,\qquad |V\rangle=(1,0,0,\ldots)^t\,.
\end{equation} 
The entries of $D$ and $E$ are of the form
$$D_{kk}=\frac{\alpha+\delta+(k-1)(\alpha\beta+2\alpha\delta+\gamma\delta)}{(\alpha+\gamma)(\beta+\delta)}\,,\quad E_{kk}=\frac{\beta+\gamma+(k-1)(\alpha\beta+2\beta\gamma+\gamma\delta)}{(\alpha+\gamma)(\beta+\delta)}\,, $$
$$D_{k,k+1}=\frac{\alpha}{\alpha+\gamma}\sqrt{k(\lambda+k)}\,,\quad E_{k,k+1}=\frac{\gamma}{\alpha+\gamma}\sqrt{k(\lambda+k)}\,,$$
$$D_{k+1,k}=\frac{\delta}{\beta+\delta}\sqrt{k(\lambda+k)}\,,\quad E_{k+1,k}=\frac{\beta}{\beta+\delta}\sqrt{k(\lambda+k)}\,,$$
with
$$
\lambda=\frac{\alpha+\beta+\gamma+\delta}{(\alpha+\gamma)(\beta+\delta)}-1\,.
$$
For low length we have checked that \eqref{eq:dehppro} agrees with our result in \eqref{eq:probs}. For this purpose it is enough to consider sufficiently large matrices $D$ and $E$ of finite size.

{
\small
\bibliographystyle{utphys2}
\bibliography{refs}

\providecommand{\href}[2]{#2}\begingroup\raggedright\begin{thebibliography}{10}

\bibitem{Sklyanin:1988yz}
E.~K. Sklyanin, ``{Boundary Conditions for Integrable Quantum Systems},''
\href{http://dx.doi.org/10.1088/0305-4470/21/10/015}{{\em J. Phys.} {\bfseries
  A21} (1988) 2375--289}.

\bibitem{Faddeev:1996iy}
L.~D. Faddeev, ``{How algebraic Bethe ansatz works for integrable model},'' in
  {\em {Relativistic gravitation and gravitational radiation. Proceedings,
  School of Physics, Les Houches, France, September 26-October 6, 1995}},
  pp.~pp. 149--219.
\newblock 1996.
\newblock
\href{http://arxiv.org/abs/hep-th/9605187}{{\ttfamily arXiv:hep-th/9605187
  [hep-th]}}.
\newblock

\bibitem{Bethe1931}
H.~Bethe, ``Zur Theorie der Metalle,''
  \href{http://dx.doi.org/10.1007/BF01341708}{{\em Zeitschrift f{\"u}r Physik}
  {\bfseries 71} no.~3, (Mar, 1931) 205--226}.

\bibitem{gaudin1983fonction}
M.~Gaudin, {\em La fonction d'onde de Bethe}.
\newblock Masson, 1983.

\bibitem{Alcaraz:1987uk}
F.~C. Alcaraz, M.~N. Barber, M.~T. Batchelor, R.~J. Baxter, and G.~R.~W.
  Quispel, ``{Surface Exponents of the Quantum XXZ, Ashkin-Teller and Potts
  Models},''
\href{http://dx.doi.org/10.1088/0305-4470/20/18/038}{{\em J. Phys.} {\bfseries
  A20} (1987) 6397}.

\bibitem{Melo:2005gut}
C.~S. Melo, G.~A.~P. Ribeiro, and M.~J. Martins, ``{Bethe ansatz for the
  XXX-$S$ chain with non-diagonal open boundaries},''
  \href{http://dx.doi.org/10.1016/j.nuclphysb.2004.12.008}{{\em Nucl. Phys.}
  {\bfseries B711} (2005) 565--603},
\href{http://arxiv.org/abs/nlin/0411038}{{\ttfamily arXiv:nlin/0411038
  [nlin.SI]}}.

\bibitem{BCR}
S.~{Belliard}, N.~{Cramp{\'e}}, and E.~{Ragoucy}, ``{Algebraic Bethe Ansatz for
  Open XXX Model with Triangular Boundary Matrices},''
  \href{http://dx.doi.org/10.1007/s11005-012-0601-6}{{\em Letters in
  Mathematical Physics} {\bfseries 103} no.~5, (May, 2013) 493--506},
  \href{http://arxiv.org/abs/1209.4269}{{\ttfamily arXiv:1209.4269 [math-ph]}}.

\bibitem{Antonio:2014qxa}
N.~Cirilo~António, N.~Manojlović, and I.~Salom, ``{Algebraic Bethe ansatz for
  the XXX chain with triangular boundaries and Gaudin model},''
  \href{http://dx.doi.org/10.1016/j.nuclphysb.2014.10.014}{{\em Nucl. Phys.}
  {\bfseries B889} (2014) 87--108},
\href{http://arxiv.org/abs/1405.7398}{{\ttfamily arXiv:1405.7398 [math-ph]}}.

\bibitem{Crampe:2011fm}
N.~Cramp{\'e} and E.~Ragoucy, ``{Generalized coordinate Bethe ansatz for non
  diagonal boundaries},''
  \href{http://dx.doi.org/10.1016/j.nuclphysb.2012.01.020}{{\em Nucl. Phys.}
  {\bfseries B858} (2012) 502--512},
\href{http://arxiv.org/abs/1105.0338}{{\ttfamily arXiv:1105.0338 [hep-th]}}.

\bibitem{schutz2000exactly}
G.~Schütz,
  \href{http://dx.doi.org/https://doi.org/10.1016/S1062-7901(01)80015-X}{``Exactly
  Solvable Models for Many-Body Systems Far from Equilibrium,''} vol.~19 of
  {\em Phase Transitions and Critical Phenomena}, pp.~1 -- 251.
\newblock Academic Press, 2001.

\bibitem{Mallick}
O.~Golinelli and K.~Mallick, ``The asymmetric simple exclusion process: an
  integrable model for non-equilibrium statistical mechanics,''
  \href{http://dx.doi.org/10.1088/0305-4470/39/41/S03}{{\em Journal of Physics
  A: Mathematical and General} {\bfseries 39} no.~41, (2006) 12679},
  \href{http://arxiv.org/abs/cond-mat/0611701}{{\ttfamily
  arXiv:cond-mat/0611701 [cond-mat.stat-mech]}}.

\bibitem{blythe2007nonequilibrium}
R.~A. {Blythe} and M.~R. {Evans}, ``{Nonequilibrium steady states of
  matrix-product form: a solver's guide},''
  \href{http://dx.doi.org/10.1088/1751-8113/40/46/R01}{{\em Journal of Physics
  A Mathematical General} {\bfseries 40} no.~46, (Nov, 2007) R333--R441},
  \href{http://arxiv.org/abs/0706.1678}{{\ttfamily arXiv:0706.1678
  [cond-mat.stat-mech]}}.

\bibitem{Derrida:1992vu}
B.~Derrida, M.~R. Evans, V.~Hakim, and V.~Pasquier, ``{Exact solution of a 1d
  asymmetric exclusion model using a matrix formulation},''
\href{http://dx.doi.org/10.1088/0305-4470/26/7/011}{{\em J. Phys.} {\bfseries
  A26} (1993) 1493--1518}.

\bibitem{Crampe:2014aoa}
N.~Cramp{\'e}, E.~Ragoucy, and M.~Vanicat, ``{Integrable approach to simple
  exclusion processes with boundaries. Review and progress},''
  \href{http://dx.doi.org/10.1088/1742-5468/2014/11/P11032}{{\em J. Stat.
  Mech.} {\bfseries 1411} no.~11, (2014) P11032},
\href{http://arxiv.org/abs/1408.5357}{{\ttfamily arXiv:1408.5357 [math-ph]}}.

\bibitem{PhysRevLett.95.240601}
J.~de~Gier and F.~H.~L. Essler, ``Bethe Ansatz Solution of the Asymmetric
  Exclusion Process with Open Boundaries,''
  \href{http://dx.doi.org/10.1103/PhysRevLett.95.240601}{{\em Phys. Rev. Lett.}
  {\bfseries 95} (Dec, 2005) 240601},
  \href{http://arxiv.org/abs/cond-mat/0508707}{{\ttfamily
  arXiv:cond-mat/0508707 [cond-mat.stat-mech]}}.

\bibitem{degier2006}
J.~{de Gier} and F.~H.~L. {Essler}, ``{Exact spectral gaps of the asymmetric
  exclusion process with open boundaries},''
  \href{http://dx.doi.org/10.1088/1742-5468/2006/12/P12011}{{\em Journal of
  Statistical Mechanics: Theory and Experiment} {\bfseries 2006} no.~12, (Dec,
  2006) 12011}, \href{http://arxiv.org/abs/cond-mat/0609645}{{\ttfamily
  arXiv:cond-mat/0609645 [cond-mat.stat-mech]}}.

\bibitem{Pimenta:2013cua}
R.~A. Pimenta and A.~Lima-Santos, ``{Algebraic Bethe ansatz for the six vertex
  model with upper triangular K-matrices},''
  \href{http://dx.doi.org/10.1088/1751-8113/46/45/455002}{{\em J. Phys.}
  {\bfseries A46} (2013) 455002},
\href{http://arxiv.org/abs/1308.4446}{{\ttfamily arXiv:1308.4446 [math-ph]}}.

\bibitem{Belliard:2014fsa}
S.~Belliard, ``{Modified algebraic Bethe ansatz for XXZ chain on the segment
  – I: Triangular cases},''
  \href{http://dx.doi.org/10.1016/j.nuclphysb.2015.01.003}{{\em Nucl. Phys.}
  {\bfseries B892} (2015) 1--20},
\href{http://arxiv.org/abs/1408.4840}{{\ttfamily arXiv:1408.4840 [math-ph]}}.

\bibitem{derrida2007non}
B.~{Derrida}, ``{Non-equilibrium steady states: fluctuations and large
  deviations of the density and of the current},''
  \href{http://dx.doi.org/10.1088/1742-5468/2007/07/P07023}{{\em Journal of
  Statistical Mechanics: Theory and Experiment} {\bfseries 2007} no.~7, (Jul,
  2007) 07023}, \href{http://arxiv.org/abs/cond-mat/0703762}{{\ttfamily
  arXiv:cond-mat/0703762 [cond-mat.stat-mech]}}.

\bibitem{Derrida_2002}
B.~Derrida, J.~L. Lebowitz, and E.~R. Speer, ``Large Deviation of the Density
  Profile in the Steady State of the Open Symmetric Simple Exclusion Process,''
  \href{http://dx.doi.org/10.1023/a:1014555927320}{{\em Journal of Statistical
  Physics} {\bfseries 107} no.~3/4, (2002) 599–634},
  \href{http://arxiv.org/abs/cond-mat/0109346}{{\ttfamily
  arXiv:cond-mat/0109346 [cond-mat]}}.

\bibitem{Derrida_2006}
B.~Derrida, J.~L. Lebowitz, and E.~R. Speer, ``Entropy of Open Lattice
  Systems,'' \href{http://dx.doi.org/10.1007/s10955-006-9160-5}{{\em Journal of
  Statistical Physics} {\bfseries 126} no.~4-5, (Jul, 2006) 1083–1108},
  \href{http://arxiv.org/abs/0704.3742}{{\ttfamily arXiv:0704.3742
  [cond-mat.stat-mech]}}.

\bibitem{Frassek2019}
R.~Frassek, C.~Giardin{\`a}, and J.~Kurchan, ``Non-compact Quantum Spin Chains
  as Integrable Stochastic Particle Processes,''
  \href{http://dx.doi.org/10.1007/s10955-019-02375-4}{{\em Journal of
  Statistical Physics} (Aug, 2019) },
\href{http://arxiv.org/abs/1904.01048}{{\ttfamily arXiv:1904.01048 [math-ph]}}.

\bibitem{Frassek:2019isa}
R.~Frassek, ``{The non-compact XXZ spin chain as stochastic particle
  process},'' \href{http://dx.doi.org/10.1088/1751-8121/ab2fb1}{{\em J. Phys.}
  {\bfseries A52} no.~33, (2019) 335202},
\href{http://arxiv.org/abs/1904.02191}{{\ttfamily arXiv:1904.02191 [math-ph]}}.

\bibitem{Kitanine:2016pvg}
N.~Kitanine, J.~M. Maillet, G.~Niccoli, and V.~Terras, ``{The open XXX spin
  chain in the SoV framework: scalar product of separate states},''
  \href{http://dx.doi.org/10.1088/1751-8121/aa6cc9}{{\em J. Phys.} {\bfseries
  A50} no.~22, (2017) 224001},
\href{http://arxiv.org/abs/1606.06917}{{\ttfamily arXiv:1606.06917 [math-ph]}}.

\bibitem{Maillet:2019hdq}
J.~M. Maillet and G.~Niccoli, ``{On Separation of Variables for Reflection
  Algebras},'' \href{http://dx.doi.org/10.1088/1742-5468/ab357a}{{\em J. Stat.
  Mech.} {\bfseries 1909} no.~9, (2019) 094020},
\href{http://arxiv.org/abs/1904.00852}{{\ttfamily arXiv:1904.00852 [math-ph]}}.

\bibitem{Crampe}
N.~Cramp{\'e}, M.~R. Evans, K.~Mallick, E.~Ragoucy, and M.~Vanicat, ``Matrix
  product solution to a 2-species TASEP with open integrable boundaries,''
  \href{http://dx.doi.org/10.1088/1751-8113/49/47/475001}{{\em Journal of
  Physics A: Mathematical and Theoretical} {\bfseries 49} no.~47, (2016)
  475001}, \href{http://arxiv.org/abs/1606.08148}{{\ttfamily arXiv:1606.08148
  [cond-mat.stat-mech]}}.

\bibitem{vanicat2017exact}
M.~Vanicat, ``Exact solution to integrable open multi-species SSEP and
  macroscopic fluctuation theory,''
  \href{http://dx.doi.org/10.1007/s10955-016-1705-7}{{\em Journal of
  Statistical Physics} {\bfseries 166} no.~5, (2017) 1129--1150},
  \href{http://arxiv.org/abs/1610.08388}{{\ttfamily arXiv:1610.08388
  [cond-mat.stat-mech]}}.

\bibitem{finn2018matrix}
C.~Finn, E.~Ragoucy, and M.~Vanicat, ``Matrix product solution to multi-species
  ASEP with open boundaries,''
  \href{http://dx.doi.org/10.1088/1742-5468/aab1b5}{{\em Journal of Statistical
  Mechanics: Theory and Experiment} {\bfseries 2018} no.~4, (2018) 043201},
  \href{http://arxiv.org/abs/1712.06809}{{\ttfamily arXiv:1712.06809
  [cond-mat.stat-mech]}}.

\bibitem{Belliard:2013aaa}
S.~Belliard and N.~Cramp{\'e}, ``{Heisenberg XXX Model with General Boundaries:
  Eigenvectors from Algebraic Bethe Ansatz},''
  \href{http://dx.doi.org/10.3842/SIGMA.2013.072}{{\em SIGMA} {\bfseries 9}
  (2013) 072},
\href{http://arxiv.org/abs/1309.6165}{{\ttfamily arXiv:1309.6165 [math-ph]}}.

\bibitem{Cao:2014aqa}
X.~Zhang, Y.-Y. Li, J.~Cao, W.-L. Yang, K.~Shi, and Y.~Wang, ``{Retrieve the
  Bethe states of quantum integrable models solved via off-diagonal Bethe
  Ansatz},'' \href{http://dx.doi.org/10.1088/1742-5468/2015/05/P05014}{{\em J.
  Stat. Mech.} {\bfseries 1505} no.~5, (2015) P05014},
\href{http://arxiv.org/abs/1407.5294}{{\ttfamily arXiv:1407.5294 [math-ph]}}.

\bibitem{Cao:2013qxa}
J.~Cao, W.-L. Yang, K.~Shi, and Y.~Wang, ``{Off-diagonal Bethe ansatz solution
  of the XXX spin-chain with arbitrary boundary conditions},''
  \href{http://dx.doi.org/10.1016/j.nuclphysb.2013.06.022}{{\em Nucl. Phys.}
  {\bfseries B875} (2013) 152--165},
\href{http://arxiv.org/abs/1306.1742}{{\ttfamily arXiv:1306.1742 [math-ph]}}.

\bibitem{Nepomechie:2013ila}
R.~I. Nepomechie, ``{An inhomogeneous T-Q equation for the open XXX chain with
  general boundary terms: completeness and arbitrary spin},''
  \href{http://dx.doi.org/10.1088/1751-8113/46/44/442002}{{\em J. Phys.}
  {\bfseries A46} (2013) 442002},
\href{http://arxiv.org/abs/1307.5049}{{\ttfamily arXiv:1307.5049 [math-ph]}}.

\bibitem{wadati}
M.~{Uchiyama}, T.~{Sasamoto}, and M.~{Wadati}, ``{Asymmetric simple exclusion
  process with open boundaries and Askey Wilson polynomials},''
  \href{http://dx.doi.org/10.1088/0305-4470/37/18/006}{{\em Journal of Physics
  A Mathematical General} {\bfseries 37} no.~18, (May, 2004) 4985--5002},
  \href{http://arxiv.org/abs/cond-mat/0312457}{{\ttfamily
  arXiv:cond-mat/0312457 [cond-mat.stat-mech]}}.

\end{thebibliography}\endgroup
}

\let\thefootnote\relax\footnotetext{\Letter$\;$ rfrassek@mpim-bonn.mpg.de}
\end{document}